\documentclass[aps,prl,reprint,amsmath,amssymb,showpacs]{revtex4-2}
\usepackage{datetime}
\usepackage{natbib}
\usepackage{changes}
\usepackage{graphicx}
\usepackage[british]{babel}

\newcommand{\ie}[0]      {i.e.}

\newcommand{\teop}        {{\mathbf T}}
\newcommand{\gteop}       {{\mathbf G}}
\newcommand{\gteopt}      {{\mathbf \Gamma}}
 \newcommand{\te}[2]      {\teop_{{#1}\to{#2}}}

\newcommand{\eqnRef}[1]{Eqn.~\ref{eqn:#1}}
\newcommand{\figRef}[1]{Fig.~\ref{fig:#1}} 

\providecommand{\energydelta}	{\Delta E_{ki}}

\providecommand{\glaubergte} 	{\gteop^{(g)}} 
\providecommand{\energygte}		{\gteop^{(e)}} 
\providecommand{\sweepgte}	{\gteop^{(s)}} 
\providecommand{\flipgte}		{\gteop^{(f)}} 
\providecommand{\latticestate}  {\mathbf{s}}
\providecommand{\nicesim}{{\raise.17ex\hbox{$\scriptstyle\mathtt{\sim}$}}}

\newif\ifletteronly
\letteronlytrue

\newif\ifnotPRL
\notPRLfalse

\begin{document}

\newcommand{\mainfile}{}

\title{Information Flow in First-Order Potts Model Phase Transition}

\author{Joshua M. Brown}
\email{joshuabrown5@acm.org}
\affiliation{School of Computing \& Mathematics, Charles Sturt University, Bathurst, NSW, Australia}

\author{Terry Bossomaier}
\affiliation{Centre for Research in Complex Systems, Charles Sturt University, Bathurst, NSW, Australia}

\author{Lionel Barnett}
\affiliation{Sackler Centre for Consciousness Science, Department of Informatics, University of Sussex, Brighton, U.K.}


\begin{abstract}
Phase transitions abound in nature and society, and, from species
extinction to stock market collapse, their prediction is of widespread
importance. In earlier work we showed that Global
Transfer Entropy, a general measure of information flow, was found to
peak away from the transition on the disordered side for the Ising
model, a canonical second-order transition~\citep{barnett13}.
Here we show that (a) global transfer entropy also peaks on the disordered side of the transition of
finite first-order transitions, \ie, those which have finite latent heat and no correlation length divergence, 
such as ecology dynamics on coral reefs~\citep{Fung11}, and (b) analysis of
information flow across state boundaries unifies both transition
orders.
We obtain the first information-theoretic result for the high-order
Potts model and the first demonstration of early warning of a  first-order transition.  The
unexpected earlier finding that global transfer entropy peaks on the
disordered side of a transition is also found for finite first-order  systems,
albeit not in the thermodynamic limit. By noting that  the interface
length of clusters in each phase is the dominant region of information
flow, we unify the information theoretic behaviour of first and second-order transitions.
\end{abstract}

\maketitle

Numerous mechanisms for predicting phase transitions exist, applied
 for example, 
from core science and engineering through biology, ecology, medicine
and finance~\citep{scheffer12}: increased variance and critical slowing
down~\citep{scheffer12}; flickering~\citep{Wang12}; and a peak in the global transfer entropy\cite{barnett13,brown20}
(\eqnRef{gte}).
Two important 
models of equilibrium transitions stand out: the Ising model~\citep{ising25}, a binary spin
system on a square lattice, where each point on the lattice has a binary
spin; and the Potts model, which
generalises Ising to spins with an arbitrary number of states, $q$, and
reduces to the Ising model for $q=2$.

\emph{Transfer entropy,} $\teop$, measures (Eqn. 2, Eqn. 4,
suppl. material) information flow from one stochastic
process, $Y$, to another, $X$---in this case the states of two
neighbouring spins over time.
\emph{Global transfer entropy}, $\gteop$,  measures the average information flow of the entire system to individual spin sites:
\begin{equation}
\gteop = \frac1N \sum_i \te{\latticestate}{s_i}\,.
\label{eqn:gte}
\end{equation}
We note however, that all information---no matter its origin in the lattice---must flow to $s_i$ via its neighbours or its own past, and thus consider only the immediate neighbourhood of each site (including $s_i$) rather than $\latticestate$ in \eqnRef{gte}~\citep{barnett13}. As with $\teop$, $\gteop \geq 0$ with $\gteop=0$ iff each site $s_i$, conditioned on its past, is independent of its neighbours.

In the Ising model~\citep{ising25}, mutual
information peaks \emph{at} the transition
between ordered and disordered phases~\citep{matsuda96,lau13}. The 
\emph{pairwise} transfer entropy~\citep{schreiber00}~(Eqn. 4, suppl. material), a
measure of information flow between spins also peaks at the transition
(suppl. material, but the  \emph{global} transfer
  entropy (\eqnRef{gte}), measuring information flow from all spins
  to any given spin, peaks  on the disordered
  side~\citep{barnett13} regardless of lattice size.

The $q$-state Potts
  model~\citep{Potts52} exhibits increasingly first-order phase transitions
  for $q>4$~\citep{Baxter73}. 
  At $q=5$ the transition is \emph{weakly first-order}, implying a
  long correlation length and low latent heat. As $q$ increases the
  correlation length decreases and the latent heat increases.
We show that as the system becomes more strongly first-order (\ie,
$q>7$) the behaviour of $\gteop$ diverges
from the second-order behaviour: in the thermodynamic limit, $\gteop$
becomes discontinuous at the transition temperature, $T_c$, peaking at
$T_c^+$. We go on to  \emph{ provide a unifying framework for both
  transitions,} based on the phase interface.

The standard Potts model comprises a
lattice of spins with periodic boundary conditions and size $N=L\times
L$, where the system state is $\latticestate=s_1,\ldots,s_N$, with
$s_i\in\{1,\ldots,q\}$. The interaction energy between two
neighbouring sites is $E_{ij}=-J\delta(s_i,s_j)$ giving the Hamiltonian $
\mathcal{H} = -J\sum_{\langle i,j \rangle}\delta(s_i, s_j)$,
where interaction strength $J=1$, $\delta(x,y)$ is the Kronecker delta
function which is one if $x=y$ and zero otherwise, and $\langle i,j
\rangle$ are all interacting pairs of sites in the system, in this case nearest cardinal neighbours. Local site
energy, $E_i$, is defined similarly, fixing site 
$i$ and summing over its four neighbours.

Overall alignment of the lattice is measured by its \emph{magnetisation}, $
M = (q\langle s_m\rangle-1)/(q-1)$~\citep{Binder81}, where $s_m$ is the \emph{mode} state and $\langle s_m \rangle = \sum \delta(s_m, s_i) / N$ is the proportion of the dominant state
over all sites, ranging from $q^{-1}$ to $1$, giving magnetisation in
the range $[0,1]$. $M$ serves as the order parameter with  order-disorder transition occurring at temperature~\citep{Okano97}
\begin{equation}
T_c = \Big[\log(1+\sqrt{q})\Big]^{-1}\,,
\label{eqn:crittemp}
\end{equation}
where the (thermodynamic) system is disordered ($M=0$) at temperatures above $T_c$ and non-zero below $T_c$. The behaviour at $T_c$ defines the transition order, where $q\leq4$ has continuous $M$ (and discontinuous $dM/dT$) giving a second-order phase transition.

Unlike the Ising model, direct simulation of the Potts models for high
$q$ is not
straightforward, since the first-order transition shows a void region
of energy space around the phase transition. For
temperatures close to the critical temperature, the energy distribution
$P(E)$ is bimodal (See suppl. material,  Fig. 1), such that single spin
update schemes, such as Glauber
dynamics, are very unlikely to enter this region, and thus fully explore the energy space. On the other hand,
cluster update schemes such as Swendsen-Wang dynamics~\citep{swendsen87} flip large groups of spins in each update-step. This can lead to an energy change in the configuration large enough to cross the energy void.
 
Thus  we estimate $\gteop$ using $10^5$ and $10^6$ Swendsen-Wang
cluster updates to approach equilibrium followed by $10^5$ Glauber sweep steps comprising $N$ spin-flip attempts per sweep. We also estimate $\gteop$ with a second regime using the \emph{density of
  states}, $d(E)$,  calculated with the Wang-Landau algorithm~\citep{Wang01}. $P(E)$ may then be calculated from
\begin{equation}
P(E) = d(E)\exp(-E/[k_bT])\,,
\label{eqn:dospe}
\end{equation}
where $E$ is the lattice energy. Since $\gteop$ depends upon two consecutive updates
  it  also depends on the temperature, which determines the
  update statistics. $\gteop$ may now be determined from its value as a function of $\gteopt(E,T)$~\citep{Wang01a}:
\begin{equation}
G(T) = \frac{\displaystyle\sum_E \gteopt(E,T) P'(E)}{\displaystyle\sum_E P'(E)}\,,
\label{eqn:dosqt}
\end{equation}
where $\gteopt$ is $\gteop$ measured at temperature, $T$, and
energy, $E$,  and $P'(E)$ is the distribution of energies, and has been rescaled for visualisation and computational reasons \footnote{Specifically, normalisation is such that $P'(E) = \exp\Big[\log[gd(E)] - E/[k_bT] - \max(\log[d(E)] - E/[k_bT])\Big]$. As the new term, $\max(\log[d(E)] - E/[k_bT])$, is constant over the summation, it cancels out such that $f(T)$ is unmodified.}.

Therefore we in fact need to determine $\gteopt(E,T)$ for varying $T$, rather than $\gteopt(E)$. Additionally, as $P(E) \to 0$ for many values, $\gteopt(E,T)$ can be measured more simply by culling energy values where $P(E)$ is sufficiently low---that is, reaching \emph{every} $E$ is unnecessary and thus $\gteopt(E,T)$ can be calculated via Glauber dynamics rather than Wang-Landau updating.

In the Swendsen-Wang and Glauber regime, which we denote $\glaubergte$, we collate statistics---the site, its neighbours and its future---for each site of the lattice after each Glauber sweep. We collate ensemble statistics for $\gteopt(E,T)$ in a similar fashion, for varying $T$ and using lattice energy $E$ prior to the Glauber sweep. We note however that the choice of $N$ spin-flip attempts for a Glauber sweep is to reduce intersample correlations compared to individual spin-flips and that each flip happens in serial rather than in parallel. While this is fine in $\glaubergte$, as Glauber dynamics maintain detailed balance, for $\gteopt(E,T)$ we collate statistics according to their specific $E$ value while also noting that $\gteop$ is a temporal quantity. Thus each spin-flip attempt during a sweep will have its own \emph{a priori} and \emph{a posteriori} states as well as its own energy value, $E'$, which may not necessarily equal $E$ and therefore statistics should ideally be collated into $\gteopt(E',T)$ rather than $\gteopt(E,T)$. Thus to explore the effect of sweep size, we employ two timescales for the density of states approach: the typical $N$ spin-flip sweep ($\sweepgte$) and the minimum, single spin-flip ($\flipgte$).

$\gteop$ is estimated via plug-in discrete  entropy histogram-based
estimators from a single realisation with settling time of $10^5$ or $10^6$ Swendsen-Wang
update-steps, followed by a measurement sequence of $10^5$ time steps (using
Glauber dynamics). Standard error is calculated by
repeating the experiment $10$ times. We optimise simulation by modifying
initialisation dependent on $T$. Realisations are initialised to
the disordered regime (\ie, each site is set, independently and uniformly, to a random state).
Experiments involving the density of states approaches are constructed
likewise, minus the superfluous (in this regime only) settling time. Settling time is unnecessary in this regime since we calculate $\gteop$ directly via \eqnRef{dosqt}, where the distribution of energies is calculated beforehand and thus any $\gteop$ measured is useful, not just those at equilibrium. 

The six dimensions (a site, its four neighbours and its future) of $q$
elements in each dimension
necessitate infeasibly many data points to accurately calculate
$\gteopt(E,T)$. 
But since transition probability of a spin-flip depends only upon the
number of spins matching the initial and final spins, rather than the
spatial configuration of the  neighbouring states, it is possible to substitute the neighbour dimensions with the  current site energy
$E_i$\footnote{An alternate approach, using the energy delta
  $\energydelta$, is incorrect as information becomes double counted
  in $\gteop$.}. This regime was  validated by applying it to
$\glaubergte$, giving $\energygte$ (shown in the supplementary
materials). 

\begin{figure*}
\includegraphics[width=\textwidth]{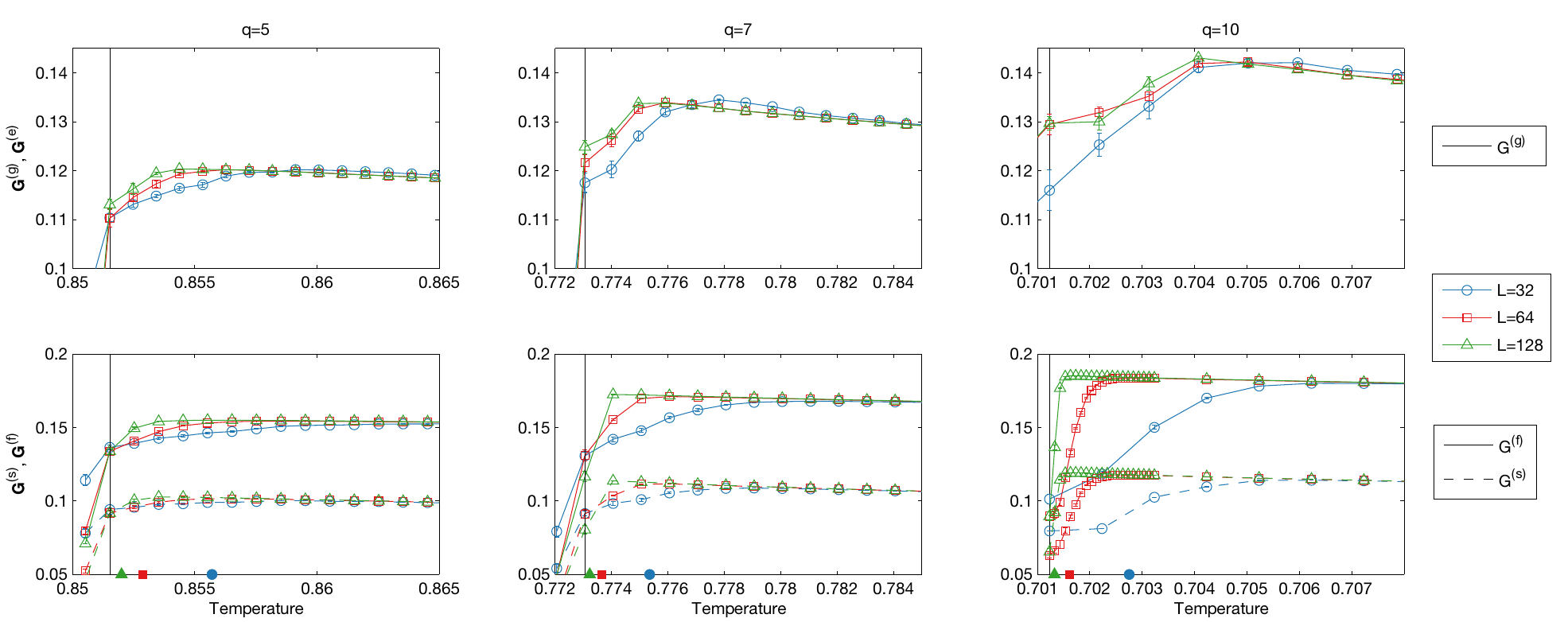}
\caption{$\gteop$ measured using three methods (top: $\glaubergte$, bottom: $\flipgte, \sweepgte$), with $q=5,7,10$ (columns) for $L=32,64,128$. $\glaubergte$ simulated for $10^5$ Swendsen-Wang update-steps, followed by $10^5$ Glauber measurement time-steps. $\flipgte, \sweepgte$ estimated using just $10^5$ Glauber measurement time-steps. Vertical lines indicate $T_c$. Filled symbols indicate ``\emph{effective}'' $T_c(L)$, the location where $P(E)$ is precisely bimodal for given $q,L$, corresponding to values found in analytical methods~\citep{Binder95}. Error bars calculated from $10$ repetitions and are smaller than symbols in some regions. Gap between $\sweepgte$ and $\flipgte$ due to extraneous data included in $\sweepgte$ (See main text).}
\label{fig:multiCompare}
\end{figure*}

Both timescales exhibit a peak in $\gteop$ on the disordered side of the
transition (\figRef{multiCompare}),  with per-sweep versus per-flip
statistics
differing by a roughly constant factor: the statistics collated for
$\sweepgte$ can be considered equivalent to those collated for
$\flipgte$ with a small amount of random noise added (\ie, those statistics collated
for $\gteopt(E,T)$ with an initial energy of $E' \neq E$, as mentioned above), thus reducing
the information flow and therefore $\sweepgte \approx c\flipgte$, with
$0<c<1$. 
  $\flipgte, \sweepgte$, exhibit
a strong shift in $\gteop$ peak as $q$ and lattice size increase,
rapidly approaching the critical temperature. Thus \emph{for the first-order
transition maximum $\gteop$ occurs at the critical temperature in the
thermodynamic limit}. Note the system displays strong
finite size effects. L=128 gets very close to the transition line
(computational limitations precluded further increases in L). The same
behaviour is visible in the top tryptich of \figRef{multiCompare}
using Swendsen-Wang updates to achieve approximate equilibrium.

Conversely, \emph{for finite systems, $\gteop$ peaks distinctly on the disordered side of $T_c$} (and the ``effective'' transition temperature $T_c(L)$~\figRef{multiCompare}~\citep{Binder95}), thus demonstrating an early warning of an impending first-order transition when approaching from the disordered side, similar to that previously demonstrated for Ising model systems~\citep{barnett13}.

\begin{figure}
\includegraphics[width=0.8\columnwidth]{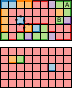}
\caption{Interfaces for $q=5$ lattice sampled from $T=T_c$ ($0.8515$) (Top) and $T=0.5$ (Bottom) where each square is a lattice site. Top: Arrows show the counter-clockwise path interface walker (for large cluster) takes around complex interactions. Labelled clusters, while sharing the same state, are disjoint, and thus have separate interfaces. Average interface length is $(34 + 3\cdot 8 + 3 \cdot 6 + 9 \cdot 4)/16=7$. Bottom: When one cluster dominates, it no longer has an ``outer'' perimeter. Average interface length is $(6+4\cdot 4)/5=4.4$.}
\label{fig:interfaceDemo}
\end{figure}

Finally, we look at a physical understanding of the behaviour of
$\gteop$.  Intuitively, information flows when neighbour states differ, hence zero information flow in ground states. This behaviour necessarily extends to clusters of states, implying information flow occurs on the boundaries, or interfaces, between clusters (See \figRef{interfaceDemo}). It seems reasonable then to assume that information flow scales with number of interfaces. However, such a maximum coincides with the zero-energy fully-disordered regime, where quite clearly $\gteop=0$. This assumption neglects the temporal nature of $\gteop$, which is disrupted at high temperature.

The average interface length is defined as:
\begin{equation}
\langle I_l \rangle = \frac{\sum_x^{N_I} I_{(x,l)}}{N_I}\,,
\label{eqn:avgInterfaceLength}
\end{equation}
where $N_I$ interface lengths are found by performing a ``\emph{turn-right walk}'' procedure, similar to~\citet{Saberi09}, on every unmarked edge between adjoining lattice sites of differing states. Edges are marked in association with an adjoining site (such that each edge is ultimately marked zero or twice). This prevents a cluster from counting its perimeter (of length $N_i$) $N_i$ separate times, but accounts for interface boundaries between clusters of two or more differing states. This also addresses clusters with two or more disjoint interfaces, \ie, a 2D doughnut.

$I(T)$ is calculated from $I(E)$ and \eqnRef{dosqt} (where $I(E)$ replaces $\gteopt(E,T)$) with the weighted Wang-Landau update scheme~\citep{Wang01}. Each $E$ value sampled at minimum $5000$ times, up to a maximum of $10000$ samples.

Remember that $\gteop$ is a measure of a site's dependence on neighbouring sites, conditioned on its own past. At high temperature, spin-flips are essentially random, choosing new states with little influence from neighbours. As temperature decreases, neighbour influence increases, leading to clusters of similar sites. We can thus approximate average influence by probability of cluster size, $p(c)$. This influence is the manifestation of information flow in the system, but only on cluster boundaries (since information flow is conditioned on its own past), leading to:
\begin{equation}
\gteop \propto \sum_c p(c)L_c\,,
\label{eqn:gteInterfaceApprox}
\end{equation}
where $L_c$ is the boundary length of cluster of size $c$. Note however that when clusters get sufficiently large---\ie, on the order of system size $L$---they no longer have an outer perimeter and are instead defined by the holes created by other clusters (\figRef{interfaceDemo}, bottom). Thus for this dominant cluster to increase in size, the internal holes must shrink and its boundary length $L_c$ actually falls. As temperature decreases, influence increases, but the available sites to transfer influence decreases, hence total information flow $\gteop$ falls.

We note that \eqnRef{gteInterfaceApprox} is essentially the \emph{average} interface length as defined in \eqnRef{avgInterfaceLength}. There should thus be some relationship between average interface length and net information flow in the lattice.

\begin{figure}
\includegraphics[width=\columnwidth]{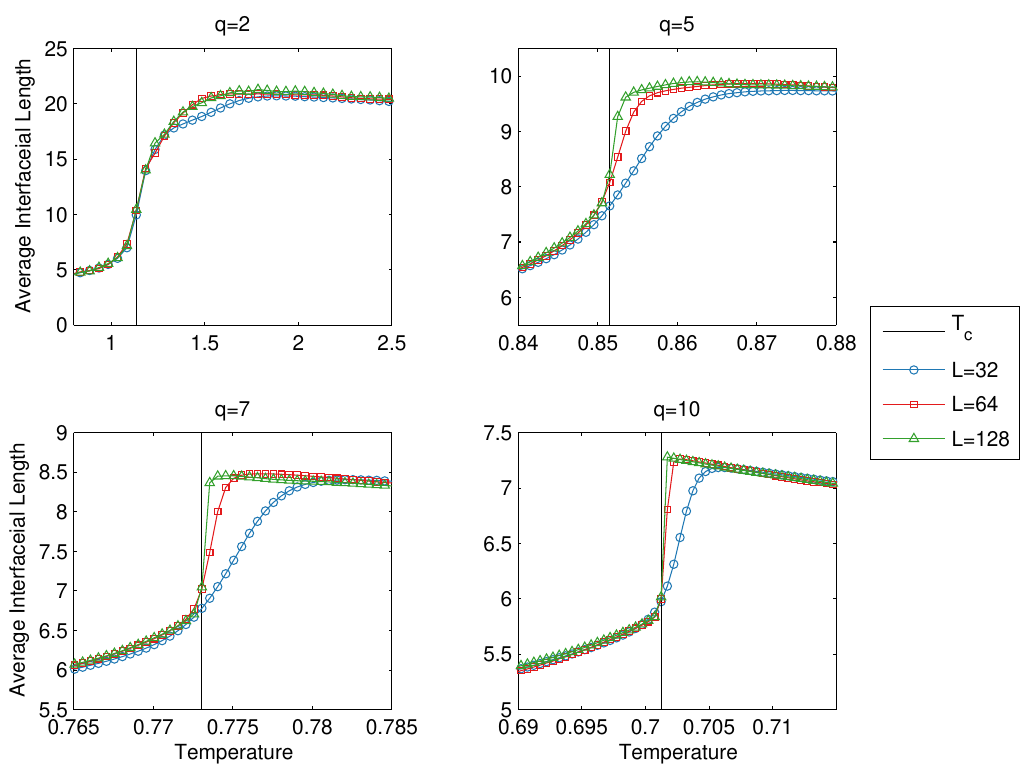}
\caption{Average interface length for systems with $q\in\{2,5,7,10\}$ for indicated lattice sizes. The behaviour in peak location mimics the behaviour of the $\gteop$ peak in all systems: the first-order cases, $q\in\{5,7,10\}$, converge to $T_c$ as the system becomes more strongly first-order (increased $q,L$), while the second-order peak $q=2$ remains stable above the phase transition. Note the factor of two difference in temperature for $q=2$ and Ising results is simply due to a slight difference in definition of site energy (\ie, $E_{ij}$), with no further side effects.}
\label{fig:interfaces}
\end{figure}

The intuitive interface model of \eqnRef{gteInterfaceApprox}, shown in \figRef{interfaces}, gives a remarkably good match to the $\gteop$ trends, peaking in the disordered regime in all cases, and converging to $T_c$ only where systems become more strongly first-order (increased $q$ and increased $L$ for $q>4$). In the $q=2$ Ising case, interface peak location remains stable at increasing lattice sizes, as does $\gteop$ peak location~\citep{barnett13}.

Thus the average interface length is a suitable theoretical justification for $\gteop$, fitting the behaviour for the first- and second-order transitions into a single unified framework.

%

\section{Acknowledgements}
The National Computing Infrastructure (NCI) facility and Intersect
NSW,  provided computing time for the simulations under project e004, with part funding under Australian Research Council Linkage Infrastructure grant LE140100002.

Joshua Brown would like to acknowledge the support of his Ph.D. program and this work from the Australian Government Research Training Program Scholarship.

Lionel Barnett's research is supported by the Dr. Mortimer and Theresa
Sackler Foundation at  the University of Sussex.

\end{document}


\newcommand{\mainfile}{}

\title{Information Flow in First-Order Potts Model Phase Transition}

\author{Joshua M. Brown}
\email{joshuabrown5@acm.org}
\affiliation{School of Computing \& Mathematics, Charles Sturt University, Bathurst, NSW, Australia}

\author{Terry Bossomaier}
\affiliation{Centre for Research in Complex Systems, Charles Sturt University, Bathurst, NSW, Australia}

\author{Lionel Barnett}
\affiliation{Sackler Centre for Consciousness Science, Department of Informatics, University of Sussex, Brighton, U.K.}

\maketitle

\section{\label{sec:suppl}Supplementary Material}

\subsection{Glauber Dynamics}
The system is updated using Glauber dynamics~\citep{Glauber63}, where site $s_i$ transitions to state $s_k$ with probability
\begin{equation}
P(s_i \to s_k) = \Big[1+e^{\energydelta/(k_bT)}\Big]^{-1}\,,
\label{eqn:transitionprob}
\end{equation}
where $T$ is the system temperature, Boltzmann's constant, $k_b$ is
taken as one, and $\energydelta$ denotes the difference in site (or
system) energy should the flip occur---\ie, $\energydelta=E_k-
E_i$. This transition probability biases spin-flips towards lower
energy states---where the ground state occurs at minimum energy when
all sites take the same state---while the system temperature inhibits
this bias, which disappears as $T \to \infty$ such that spins flip to
random states with probability $0.5$. Glauber dynamics satisfy
detailed balance~\citep{VanKampen92} and thus yield the thermal
equilibrium probabilities at stationarity. However, close to the phase
transition, it can take a large number of time steps to reach
equilibrium, thus a faster, cluster update, Swendsen-Wang if used. To
calculate the information theory quantities a statistics collection
series of Glauber updates is carried out.

\subsection{Transfer Entropy}
Transfer entropy measures information flow from one stochastic process, $Y$, to another, $X$---in this case the states of two neighbouring spins over time. It is a non-negative quantity, reaching zero iff process $X$, conditioned on its own past, is independent of the past of $Y$. Positive values indicate a statistical dependency---a reduction in uncertainty---of $X$ given knowledge of the past of $Y$. Transfer entropy is given by the time-lagged mutual information, conditioned on the past of $X$:

\begin{align}
\te{Y}{X} &= \cmutinf{X_t}{Y_{t-1}}{X_{t-1}}\,,\label{eqn:teMI}\\
&=\centro{X_t}{X_{t-1}} - \centro{X_t}{X_{t-1},Y_{t-1}} \label{eqn:teCentro}\,,
\end{align}
where we use a single-step time-lag and the pairwise transfer entropy is simply the average transfer entropy over all interacting sites:
\begin{equation}
\teop_{pw} = \frac{1}{N}\sum_{\langle i,j \rangle}\te{s_j}{s_i}\,.
\label{eqn:tpw}
\end{equation}

\subsection{Energy Space}
\begin{figure}
\includegraphics[width=\columnwidth]{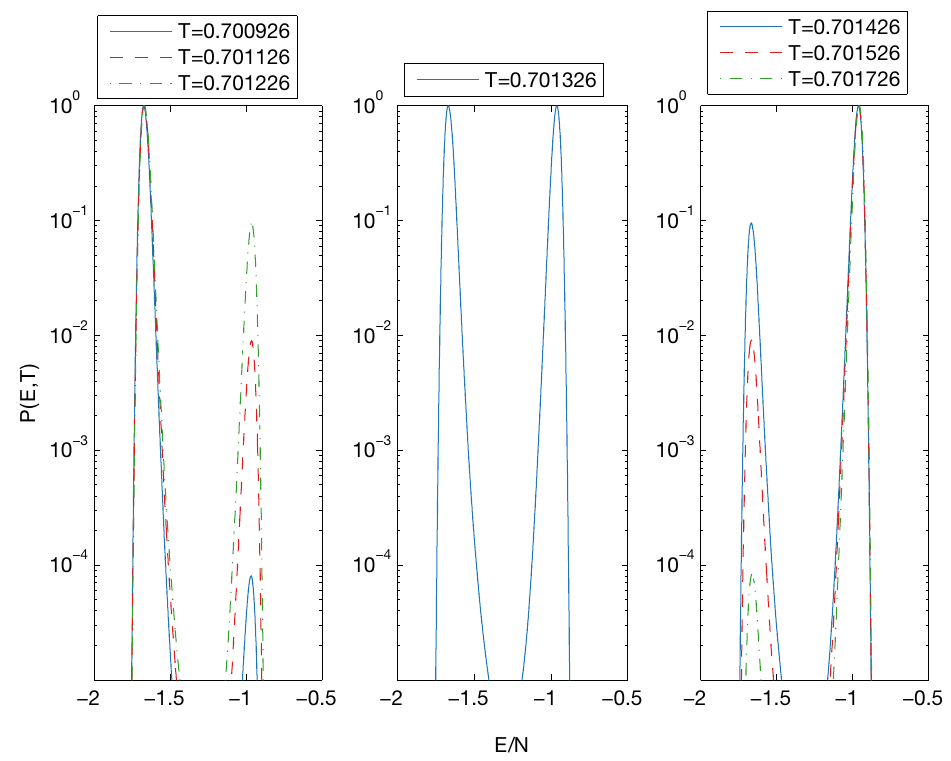}
\caption{$P(E)$ of Potts states for $q=10, L=128$. Left: $P(E)$ as $T\to T_c^-$, demonstrating emergence of right (disordered) peak. Middle: The location of the ``effective'' transition, $T_c(L)$, defined by equal height peaks. Right: $P(E)$ as $T$ moves away from $T_c(L)$, showing dissolution of ordered peak. Note that in the thermodynamic limit, each peak only exists in its relevant regime and emergence of bimodal peaks is instantaneous at $T_c$.}
\label{fig:bimodal}
\end{figure}

First-order transition show a void region
of energy space around the phase transition, such that single spin 
update schemes, such as Glauber
dynamics, are very unlikely to enter this region. In fact, for
temperatures close to the critical temperature, the energy distribution
$P(E)$ is bimodal (See \figRef{bimodal}---note a scaling factor is
introduced such that peak maximum is one). As $q$ decreases, the peaks
shift closer together until they merge into a unimodal peak at $q=4$
(characteristic of a second-order transition). The valley between
peaks is shallower for given lattice size at lower $q$, thus $q=5$ is
considered weakly first-order, while $q=10$ is strongly
first-order. As $L$ increases (with constant $q$) the valley deepens,
making simulation for large lattices, particularly at $q=10$,
increasingly difficult. 

\subsection{Neighbourhood Compression}
As discussed in the main paper, capturing data for $\gteop$ requires a 6D histogram---one dimension
for each neighbour, current spin, and future spin---which requires
$q^{12}$ data points for effective estimation (using the heuristic
$B=\sqrt{N}$~\citep{Venables02}, where $B=q$ is the number of bins in each dimension of the histogram). This volume of data is difficult, yet achievable, for a single
histogram (as in $\glaubergte$), but completely infeasible for $E$
histograms, to calculate $\Gamma(E,t)$ as required in
$\sweepgte$ and $\flipgte$. To compress the histogram.
 we note that the transition probability depends only upon the number
of spins matching the initial and final spins, rather than the exact
neighbouring states. An intuitive  approach replaces the neighbour
dimensions with the energy delta term, $\energydelta$, as this should
encode all transition information. This approach is incorrect however
as it incorporates information about the future state directly into
the conditioned variables in the second term of
\eqnRef{teCentro}---that is, $Y_{t-1}$ incorrectly becomes some function of $X_t$.

\begin{figure*}
\includegraphics[width=\textwidth]{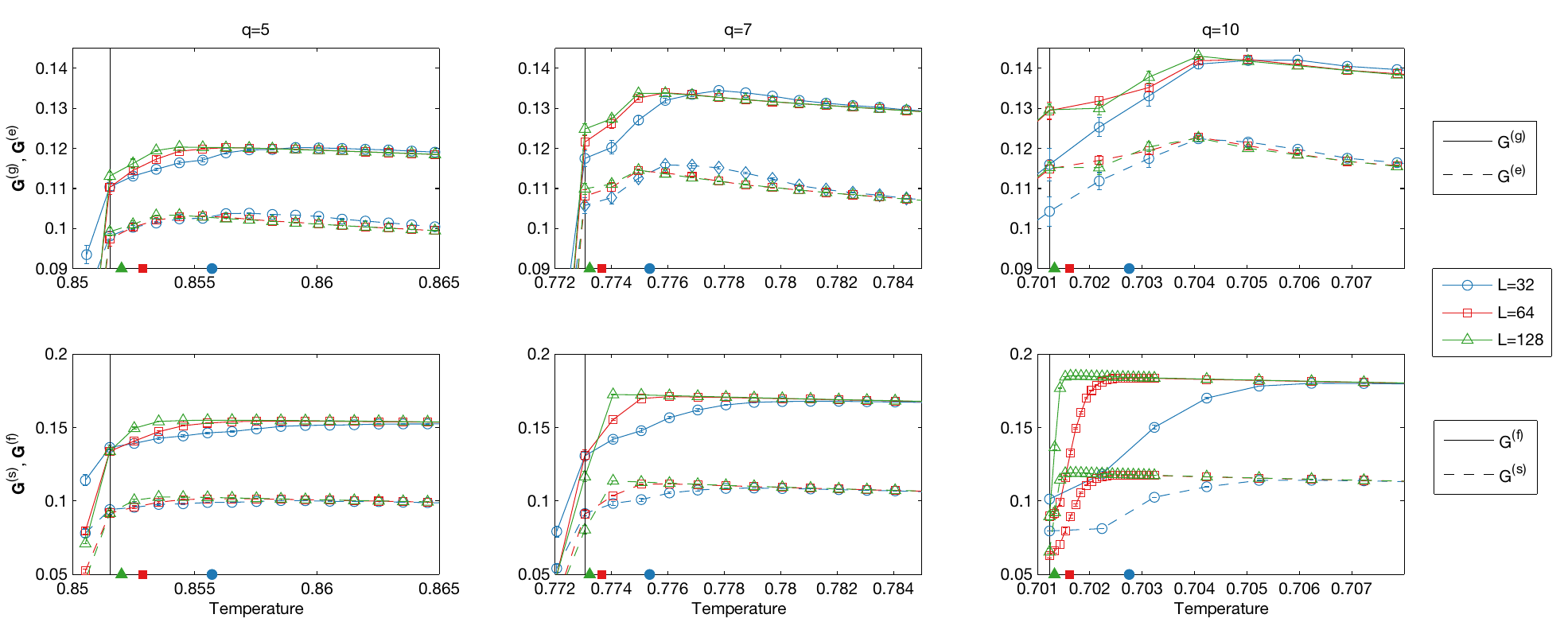}
\caption{Figure 1 (main text) repeated with validation from $\energygte$. Note that $\energygte$ tracks $\glaubergte$ with a constant reduction due to information lost in the compression algorithm.}
\label{fig:multiCompareEnergy}
\end{figure*}

Thus we encode just the current site energy, $E_i$, although not
without trade-off: removal of neighbour details leads to consistent
reduction in total available information (See
\figRef{multiCompareEnergy}, top row). This approach was  validated
with an alternative reduction with consistent results---where the
binary function, $\delta(s_i, s_j)$, is used for each neighbour. These
approaches give significant reductions in data
requirements---$(5q^2)^2$ and $(2^4q^2)^2$ respectively. The former
approach was employed as it requires fewer bins, and thus data points, without effect on the result.

\subsection{Limiting Behaviour}
\begin{figure}
\includegraphics[width=\columnwidth]{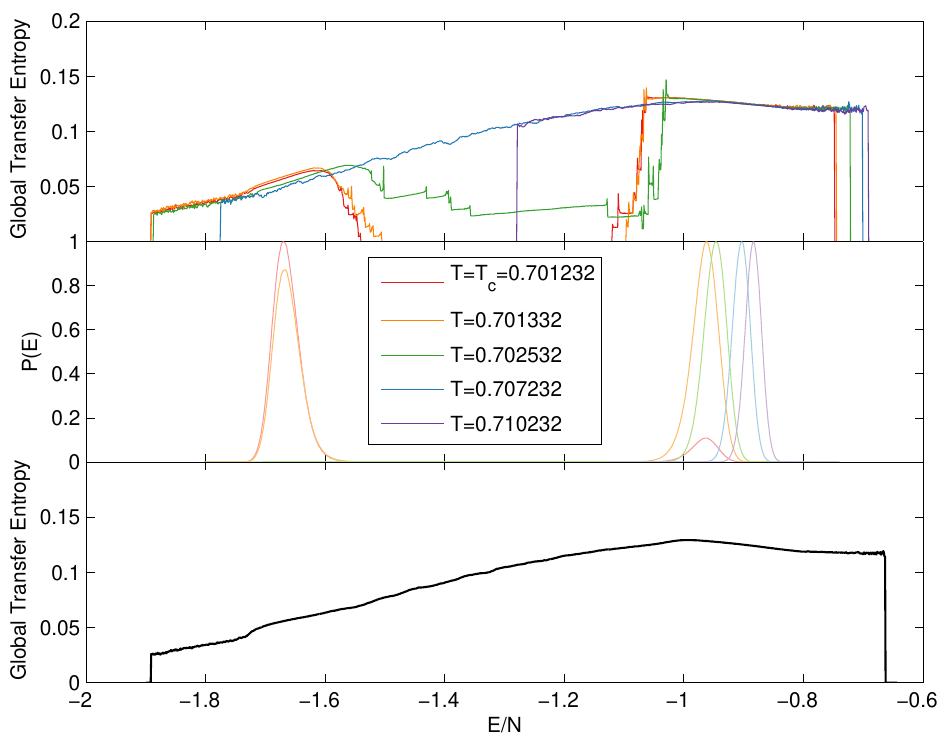}
\caption{$\flipgte(E,T)$ (top) for selected temperatures at and above $T_c$ with $P(E)$ (middle) for $q=10, L=128$ with 10 realisations, half initialised to random ground states and half to disordered states. For temperatures near $T_c$, we observe a valley in $\flipgte(E,T)$ as in $P(E)$, where realisations are unable to traverse. As $T$ increases, central energy values are reachable, with lower energy values becoming unreachable---note that $\flipgte(E,T)$ drops to zero at $E/N \approx -1.75, -1.3$ for $T=0.707, 0.710$, respectively. Bottom shows $\flipgte(E,T)$ average over $T$ where each $\flipgte(E)$ is scaled with respect to frequency over $E$.}
\label{fig:gteet}
\end{figure}

The limiting behaviour of $\gteop$ can be determined via closer analysis of $\gteopt(E,T)$. Figure~\ref{fig:gteet} shows $\flipgte(E,T)$ and $P(E)$ at $q=10, L=128$ for selected temperatures. As with the above regimes, realisations are initialised evenly between random ground states and disordered states. We can observe the valley in $\flipgte(E,T)$ which realisations are unable to traverse, noting that for $T=0.702, 0.707$, while no ordered $P(E)$ peak exists, $\flipgte(E,T)$ is non-zero due to initialisation regime. As $T$ increases, the system is able to move through this region of energy space, until high enough temperatures are reached such that lower energies become impossible, with the ground state realisations very rapidly becoming disordered.

Consider now the extreme energies (effectively temperatures) in \figRef{gteet}. On the disordered end ($E/N=-1$), we can see that $\flipgte(E,T)$ peaks below the disordered $P(E)$ peaks, and steadily decreases at higher energies (and thus temperatures), consistent with the expectation of reduced $\gteop$ as spins become increasingly independent. Similarly, as $T\to 0$, low energy $\flipgte(E,T)$ goes to zero as well: conditioned on its own past, $s_i$ becomes independent of its neighbourhood---the neighbourhood adds no additional information to knowing the past of $s_i$---as expected.

The observation of high energy $\flipgte(E,T)$ peaking earlier than $P(E)$, in the void region, also resolves the limiting behaviour near $T_c$. Specifically, when moving towards $T_c$ (and thus $P(E)$ peaks at progressively lower $E/N$) from high temperatures $\flipgte(E,T)$ is always increasing. Therefore in the thermodynamic limit, where $P(E)$ is unimodal until precisely $T_c$, $\gteop$ will increase towards $T_c$. The peaks appearing in Fig. 1 (main text) away from $T_c$ are then due to the finite size effect of bimodal $P(E)$ away from $T_c$ where low $\gteop$ ordered regimes are incorrectly sampled. Furthermore, in the limit at $T_c$, a system will be either ordered or disordered with valley $P(E)=0$, and consequently $\gteop$ will be undefined at $T_c$. 

%